\def\draftversion{false}

\RequirePackage{ifthen}
\ifthenelse{\equal{\draftversion}{true}}{
  \documentclass[aps,prb,10pt,galley,amsmath,amssymb,showpacs,
       longbibliography,superscriptaddress]{revtex4-1}
}{ 
  \documentclass[aps,prb,10pt,twocolumn,amsmath,amssymb,showpacs,
       longbibliography,superscriptaddress]{revtex4-1}
}

\usepackage{graphicx}
\usepackage[usenames]{color} 
\usepackage{bm} 
\usepackage{soul}
\usepackage{hyperref} 
\hypersetup{		  
    colorlinks=true,
    linkcolor=blue,
    filecolor=blue,      
    urlcolor=blue,
    citecolor=blue
}

\ifthenelse{\equal{\draftversion}{true}}{
  \marginparwidth 2.7in
  \marginparsep 0.5in
  \newcounter{comm} 
  \def\commnext{\stepcounter{comm}}
  \def\commtext{{\bf\color{blue}[\arabic{comm}]}}
  \def\commmar{{\bf\color{blue}[\arabic{comm}]}}
  \def\dvm#1{\commnext\marginpar{\small DV\commmar: #1}\commtext}
  \def\nvm#1{\commnext\marginpar{\small NV\commmar: #1}\commtext}
  \def\mlab#1{\marginpar{\small\bf #1}}
  
}{
  \def\dvm#1{}
  \def\nvm#1{}
  \def\mlab#1{}
  
}

\def\Changes#1{\textcolor{black}{#1}}

\def\beq{\begin{equation}}
\def\eeq{\end{equation}}
\newcommand{\equ}[1]{Eq.~(\ref{eq:#1})}

\def\bise{Bi$_2$Se$_3$}

\newcommand{\ket}[1]{\vert#1\rangle}
\newcommand{\bra}[1]{\langle#1\vert}

\renewcommand{\k}{{\bf k}}     

\newcommand{\nhat}{{\hat{\bf n}}}

\newcommand{\T}{{\cal T}}
\newcommand{\I}{{\cal I}}

\def\im{\mathrm{Im}\,}
\newcommand{\Z}{\mathbb{Z}}
\newcommand{\zt}{\Z_2}

\newcommand{\Cp}{C^{\hspace{0.5pt}\prime}}

\def\eee{\!=\!}

\begin{document}


\title{Surfaces of axion insulators}

\author{Nicodemos Varnava}
\email{nvarnava@physics.rutgers.edu}
\affiliation{
Department of Physics \& Astronomy, Rutgers University,
Piscataway, New Jersey 08854, USA}

\author{David Vanderbilt}
\affiliation{
Department of Physics \& Astronomy, Rutgers University,
Piscataway, New Jersey 08854, USA}

\date{\today}
\begin{abstract}
Axion insulators are magnetic topological insulators in which the non-trivial $\zt$ index is protected by inversion symmetry instead of time-reversal symmetry. The naturally gapped surfaces of axion insulators give rise to a half-quantized surface anomalous Hall conductivity (AHC), but the sign of the surface AHC cannot be determined from topological arguments. In this paper, we consider topological phenomena at the surface of an axion insulator. To be explicit, we construct a minimal tight-binding model on the pyrochlore lattice and investigate the all-in-all-out (AIAO) and ferromagnetic (FM) spin configurations. We also implement a recently proposed approach for calculating the surface AHC directly, which allows us to explore how the interplay between surface termination and magnetic ordering determines the sign of the half-quantized surface AHC. In the case of AIAO ordering, we show that it is possible to construct a topological state with no protected metallic states on boundaries of any dimension (surfaces, hinges, or corners), although chiral hinge modes do occur for many surface configurations. In the FM case, rotation of the magnetization by an external field offers promising means of control of chiral hinge modes, which can also appear on surface steps or where bulk domain walls emerge at the surface.

\end{abstract}

\maketitle


\section{Introduction}
\label{sec:intro}

Topological insulators are classified according to the rule that two states belong to the same class if one can be gradually converted into the other without gap closure.  Such a topological classification represents a paradigm shift in the way we understand phases of matter.  One of the most widely recognized aspects of the theory of topological states is the ``bulk-boundary correspondence,'' which refers to the notion that the boundary between two insulating states with different topological indices, or between a nontrivial topological state and the vacuum, must display metallic states at the boundary.  These are robust in the sense that they cannot be removed by any modification at the boundary or in the bulk, as long as the bulk topological state remains unchanged. This is the case for the chiral states on the surface of a Chern insulator, and for the helical Dirac states on all surfaces of a strong topological insulator (TI) when time-reversal ($\T$) symmetry is preserved.

What is not so widely appreciated is that the bulk-boundary correspondence is not always enforced. In particular, if the bulk topological classification relies on the presence of some symmetries, then the bulk-boundary correspondence typically comes into play only if the surface also respects the same symmetries.  For example, the $\zt$ topology of a strong TI is protected by $\T$; if one breaks $\T$ symmetry at the surface, as for example by proximity to a ferromagnetic coating material, the surface states are no longer guaranteed to be present.

In some cases, however, topological behavior can still be left behind at the boundary.  In the case of a strong TI, an insulating surface resulting from local breaking of $\T$ is guaranteed to exhibit a half-quantized anomalous Hall conductivity (AHC) of $e^2/2h$,\citep{qi-prb08} as was recently verified experimentally.\citep{chang-s13,wu-s16,xiao-prl18} This is connected with the fact that the axion phase angle $\theta$ characterizing the formal Chern-Simons magnetoelectric coupling\cite{qi-prb08,essin-prl09+e} takes the values $\theta\!=\!\pi$ for strong TIs and $\theta\!=\!0$ otherwise, corresponding to the ``axion $\zt$'' classification.  A half-integer surface AHC is then required at any insulating boundary separating $\T$-invariant phases with different axion $\zt$ indices.\cite{qi-prb08,essin-prl09+e}

Moreover, Essin \textit{et al.}~\citep{essin-prl09+e} have shown that one does not have to start with a strong TI and break $\T$ symmetry to acquire a half-quantum surface AHC. Indeed, the axion $\zt$ index is also well defined in the absence of $\T$,  as long as inversion symmetry $\I$ is present. Since inversion is always broken at the topological boundary, the half-quantized surface AHC can be observed without the need for any special treatment of the surface.  These magnetic materials, which are usually referred to as ``axion insulators,'' are intriguing candidates for theoretical and experimental exploration. Recently, experiments\citep{xiao-prl18,mogi-s17} approached the design of axion insulators by means of heterostructures comprised of magnetically doped and undoped thin films of TIs. Unfortunately experimental attempts to identify bulk axion insulators have yet to be successful.

In this paper we explore various topological phenomena at the surface of an axion insulator, many of which cannot be obtained from a strong TI with broken $\T$ on its surface. In doing so we hope to raise the awareness of the community regarding this promising class of materials, in the expectation that they may soon be realized and their unique surface properties explored. We investigate the conditions under which the surface magnetic point group forces the surface AHC to vanish, and when it does not, what factors decide the sign of the surface AHC. We find that the answers depend on the bulk and surface symmetry, the specific surface termination, and the magnetic configuration of the surface, yielding a plethora of possibilities to explore and manipulate.

Specifically, we carry out these investigations in the context of a model for the $R_2$Ir$_2$O$_7$ pyrochlore iridates ($R$ is typically a rare earth).  Many members of this class exhibit a low-temperature insulating phase with AIAO antiferromagnetic order, although some also display FM or other magnetic orderings.\cite{tomiyasu-psj12,sagayama-prb13,disseler-prb14,zhu-prb14,donnerer-prl16} The AIAO pyrochlores were proposed early on by Wan \textit{et al.}\citep{wan-prb11} to be in an axion-insulator phase for some range of values of the on-site Hubbard $U$.  However, they become trivial insulators for sufficiently large values of $U$, and later studies have led to a consensus that these systems are in fact in the trivial phase (see, e.g., Ref.~\onlinecite{zhang-prl17}).

Nevertheless, motivated by the interest in this material system, we have adopted a minimal tight-binding model of the magnetic sites in a pyrochlore iridate, but with weakened magnetic exchange splitting, as a platform for investigating hypothetical axion-insulator materials.  Our model consists of spinors on the Ir sites; these form a simple pyrochlore lattice, and therefore may also be relevant to some spinels.  The Hamiltonian includes on-site Zeeman splitting terms representing the effects of the magnetic order, as well as spin-dependent and independent nearest-neighbor hoppings.  The phase diagram of our model exhibits a trivial insulator phase for large Zeeman splitting, and then passes through a Weyl semimetal phase before entering an axion-insulator phase as the Zeeman splitting is reduced. We explore the behavior of the bulk and surface properties of our model primarily in this axion-insulator region of the phase diagram, focusing mainly on the case of AIAO magnetic order where the symmetry is highest, but considering FM and other orderings as well. In particular, we make use of a recently introduced numerical approach,\cite{rauch-prb18} to derive an expression that allows for an explicit calculation of the magnitude and sign of the surface AHC. In turn, this tool is used to demonstrate the half-quantized nature of the surface AHC and to reveal how the surface structural and magnetic configuration determines its sign.

The paper is organized as follows.  In Sec.~\ref{sec:axion-intro}, we establish the theoretical ideas that will be invoked in this work. Namely, the connection between the axion coupling and the surface AHC is discussed in Sec.~\ref{sec:axion}, while the role of symmetries is discussed in Sec.~\ref{sec:symm}. In Sec.~\ref{sec:CSAHC} we derive an expression for the ``partial Chern number," which will be used later to calculate the surface AHC. In Sec.~\ref{sec:pyro} we motivate and introduce our tight-binding model for an axion insulator on the pyrochlore lattice.

Section~\ref{sec:results}, which presents the results, is divided into two subsections. In the first we discuss the case of the AIAO spin configuration in detail, while in the second we briefly consider the FM configuration. There are subsections devoted to the bulk energy bands, the phase diagram, the surface states, the surface AHC and the Wannier bands.

In Sec.~\ref{sec:surfacephenomena} we discuss various phenomena on the surfaces of axion insulators. We first consider an FM axion insulator and explore what happens as we rotate the magnetization about an axis perpendicular or parallel to the surface, then investigate the consequences of the atomic layer sensitivity of the surface AHC in the AIAO configuration, and end with a consideration of crystallite geometries in the FM configuration. Section \ref{sec:summary} contains a brief summary and conclusions.

\section{Axion coupling and surface AHC}
\label{sec:axion-intro}
One of the unique and exciting aspects of topological phases of matter is that they can give rise to quantized physical quantities. For example, the formal polarization $P$ of a 1D insulator is either 0 or $e/2$ when the system has inversion symmetry, while the isotropic magnetoelectric coupling $\alpha_{\text{iso}}$ is 0 or $e^2/2h$ in 3D systems with either time-reversal or inversion symmetry. Perhaps surprisingly, these bulk quantities determine (up to a quantum) certain surface responses of these material systems.
In the following we explore how the interplay between topological quantities and symmetries give rise to quantized surface responses, in the context of axion insulators.

\subsection{Berry phase and axion coupling}
\label{sec:axion}

There is a nice analogy between the Berry phase $\phi$ in 1D and the axion coupling $\theta$ in 3D, that reveals their deep connection. Starting from the integral expression for the first Chern number in 2D and through a dimensional reduction procedure,~\citep{qi-prb08} one can derive the well known integral expression for the Berry phase
\begin{equation}\label{eq:phi}
\phi = \int_{BZ} dk \, \text{Tr}[\mathcal{A}]
\end{equation}
where $\mathcal{A} = \mathcal{A}^{nm}(k)$ is the 1D version of the Berry connection $\mathcal{A}_j^{nm}(\k) = \bra{u_{n\k}}i\partial_{k_j}\ket{u_{m\k}}$. A physical interpretation for \equ{phi} came from King-Smith and Vanderbilt,\citep{kingsmith-prb93} who showed that the Berry phase and polarization in 1D are one and the same thing, 
\begin{equation}
P = -e \, \frac{\phi}{2\pi} \, .
\end{equation}
Note that $\phi$ is not quantized and it is only well defined modulo $2\pi$, as it is evident from a gauge transformation of the cell-periodic Bloch functions $\ket{u_{n\k}} \rightarrow e^{-i\beta(\k)}\ket{u_{n\k}}$. The $2\pi$ ambiguity explains why the angle variable $\phi$ is used and why bulk polarization is well defined modulo $e$, the quantum of charge. Finally, King-Smith and Vanderbilt showed that for a finite 1D insulator there will be a charge accumulation on the surface given by
\begin{equation}
Q^{\text{surf}} = P \ \text{mod} \ e \, .
\end{equation}

We can follow the same recipe starting from the second Chern number. This is defined in four dimensions such that the dimensional reduction procedure returns a 3D expression, the Chern-Simons axion coupling~\cite{qi-prb08}
\begin{equation}\label{eq:axioncoup}
\theta = -\frac{1}{4\pi}\int_{BZ} d^3k \, \epsilon^{\alpha \beta \gamma} \text{Tr} \Big[ \mathcal{A}_\alpha \partial_\beta \mathcal{A}_\gamma -i\frac{2}{3} \mathcal{A}_\alpha \mathcal{A}_\beta \mathcal{A}_\gamma \Big] \, .
\end{equation}
As in the 1D case, a gauge transformation can cause $\theta \rightarrow \theta + 2\pi n $, $n\in \Z$, showing that it is only well defined modulo $2\pi$. Now the physical significance of this quantity is that it describes an isotropic contribution to the magnetoelectric tensor\citep{essin-prl09+e,hasan-rmp10} $\alpha_{ij} = ( \partial P_i/\partial B_j )_E = ( \partial M_i/\partial E_j )_B$. That is, an electric field will induce a parallel magnetization with a constant of proportionality that is given by
\begin{equation}\label{eq:isotr}
\alpha_{\text{iso}} = \frac{e^2}{h}\frac{\theta}{2\pi} \, ,
\end{equation}
which shows that $\alpha_{\text{iso}}$ is only well defined modulo $e^2/h$, the quantum of conductance. Continuing with our analogy, the surfaces of a 3D insulator will exhibit a surface AHC given by~\citep{coh-prb13}
\begin{equation}\label{eq:SAHC}
\sigma^{\text{surf}}_{\text{AHC}} = -\frac{e^2}{h}\frac{\theta}{2\pi} \ \text{mod} \ e^2/h \, ,
\end{equation}
as long as the surface is insulating. In fact for a metallic surface, there is another contribution from the metallic states at the Fermi energy, so that the total is~\citep{coh-prb11} 
\begin{equation}\label{eq:isotrfull}
\sigma^{\text{surf}}_{\text{AHC}} = -\frac{e^2}{h}\frac{\theta-\phi}{2\pi} \ \text{mod} \ e^2/h \, .
\end{equation}
The extra $\phi$ term is the Berry phase taken along the surface Fermi loop.  The fact that the surface AHC cannot be determined just from bulk considerations, i.e., from the knowledge of the axion coupling, can be understood as follows. Consider a block of a material with some initial $\alpha_{\text{iso}}$ which we wrap with a 2D quantum anomalous Hall insulator with Chern index $C=1$. As long as we do this without closing the surface gap, we are able to increase $\sigma^{\text{surf}}_{\text{AHC}}$ by $e^2/h$ without altering the bulk, and therefore without altering the axion coupling $\theta$.
\subsection{Symmetries}
\label{sec:symm}
The presence of $\I$ or $\T$ symmetry can restrict the Berry phase or the axion coupling in a profound way. For example, the Berry phase of a 1D insulator can only take the values $\phi = 0$ or $\pi$ when $\I$ is a good symmetry, because $\phi$ is odd under $\I$ and only these two values respect this restriction. On the other hand, $\phi$ is even under $\T$, so the presence of $\T$ enforces no restriction on $\phi$. In contrast, either $\I$ or $\T$ will reverse the sign of $\theta$, and since $\theta$ is also only well-defined modulo $2\pi$, the presence of either operator will quantize $\theta$ to be 0 (trivial) or $\pi$ (topological). The set of two possible value of $\theta$ corresponds to a topological ``axion $\zt$'' classification. Note that irrespective of the bulk topological state, the presence of $\T$ in the surface region requires $\sigma^{\text{surf}}_{\text{AHC}}$ to vanish there, although possibly by forcing the surface to be metallic.

In the usual usage, a ``strong TI'' is one for which a nontrivial $\zt$ index is protected by $\T$. In that case, $\sigma^{\text{surf}}_{\text{AHC}} = 0$ and $\theta=\pi$, so Eq.~($\ref{eq:isotrfull}$) tells us that $\phi=\pi$ on every surface. This in turn means that every surface is metallic, with an odd number of Dirac cones. Material systems in which $\T$ is absent and the nontrivial $\theta\eee\pi$ state is protected by $\I$ are referred to as ``axion insulators''. Axion insulators are remarkable in that their insulating surfaces exhibit a half-quantized surface AHC. This is easily seen from Eq.~($\ref{eq:isotrfull}$) by noting that $\phi=0$ for an insulating surface. Note that $\I$ symmetry is also present in many strong TIs such as \bise, but the term ``axion insulator'' is not typically applied to such materials unless the $\T$ symmetry is broken in some way, as for example by the application of an external magnetic field.

More generally, we can observe that $\theta$ is quantized to be 0 or $\pi$ by the presence of any operation in the magnetic point group that is either (i) a proper rotation (possibly the identity) composed with time reversal, or (ii) an improper rotation not composed with time reversal.\cite{fang-prb12} In case (i) the operation may be $\T$ itself or any of the time-reversed $n$-fold proper rotations $\Cp_n$ ($n$ = 2, 3, 4, or 6 where the prime denotes composition with $\T$).  Case (ii) pertains to mirror operations, improper rotations $S_n$, and inversion itself.  This observation follows since $\theta$ is invariant under any proper rotation, and any of the operations associated with case (i) or (ii) can be related to $\T$ or $\I$ respectively by a proper rotation.  Any insulator whose magnetic point group contains at least one of the above operations has an axion $\zt$ classification, and if it is topological ($\theta\eee\pi$) then we can regard it as an axion insulator in a generalized sense.

In the remainder of the present work we consider only the case that $\I$ itself is a symmetry, so that the system is an axion insulator in the widely accepted sense. Nevertheless, other symmetries may be present as well, and may play an important role in determining certain features of the bulk and surface band structure and the surface AHC, as will be illustrated in the context of our tight-binding model system below.

In particular, we shall be concerned with the symmetries present at a surface facet with unit normal $\nhat$.  The relevant surface magnetic point group is composed only of operations that map $\nhat$ onto itself, and whose corresponding space-group operations do not involve any fractional translation along $\nhat$. That is, screw and glide-mirror operations with fractional translations normal to the surface are excluded.  The surface AHC transforms like the component of a magnetic field or magnetization normal to the surface, and therefore it must vanish if the surface magnetic point group contains:
(i) any mirror operation $M_v$ (not $\T$-reversed and with
mirror plane normal to the surface); or
(ii) any $\T$-reversed proper rotation about $\nhat$ (i.e.,
$C'_2$, $C'_3$, $C'_4$, or $C'_6$).

\subsection{Calculation of the surface anomalous Hall conductivity}
\label{sec:CSAHC}
For the calculation of the surface anomalous Hall conductivity (AHC),
we implement a recently proposed approach\cite{rauch-prb18}
based in part on previous
developments\cite{essin-prl09+e,bianco-prb11,marrazzo-prb17}
showing that the Chern-Simons (CS) contribution to the AHC can be
expressed as a local, real space property. In this framework, one
defines a local Chern-number density
\begin{equation} \label{eq:local_surf}
\boldsymbol{C}(\boldsymbol{r}) = -2\pi\, \im\bra{\boldsymbol{r}}
  P\boldsymbol{r}Q \times Q\boldsymbol{r}P \ket{\boldsymbol{r}}
\end{equation}
from which the local CS contribution to the AHC can be obtained via
$\boldsymbol{\sigma}^{\rm CS}(\boldsymbol{r}) = (e^2/h)
\boldsymbol{C}(\boldsymbol{r})$.
Here $P$ and $Q$ are
the projection operators onto the occupied and unoccupied subspaces,
respectively. In contrast with the usual reciprocal-space integral
expression, the real space one can be used to study bounded
systems such as finite crystallines or surface slabs. This is
exactly what we need to calculate the surface AHC and determine its sign.

We apply this theory in the context of a surface slab geometry.
The unit cell is small in the in-plane direction, with primitive
lattice vectors $\boldsymbol{a}_1$ and $\boldsymbol{a}_2$
and cell area $A=|\boldsymbol{a}_1\times\boldsymbol{a}_2|$,
but it extends through the entire thickness of the slab in the
vertical $z$ direction, including top and bottom surfaces.
The Hamiltonian eigenstates in the valence ($v$)
and conduction ($c$) bands are
$\psi_{v\boldsymbol{k}}(\boldsymbol{r})$ and
$\psi_{c\boldsymbol{k}}(\boldsymbol{r})$, respectively, with
$\boldsymbol{k} = (k_x, k_y)$.
The valence and conduction projectors are then
$P=(1/N_{\boldsymbol{k}}) \sum_{v\boldsymbol{k}}
\ket{\psi_{v\boldsymbol{k}}} \bra{\psi_{v\boldsymbol{k}}}$ and
$Q = (1/N_{\boldsymbol{k}}) \sum_{c\boldsymbol{k}}
\ket{\psi_{c\boldsymbol{k}}} \bra{\psi_{c\boldsymbol{k}}}$,
where $N_{\boldsymbol{k}}$ is
the number of $\boldsymbol{k}$-points in the 2D Brillouin
zone mesh.
Plugging these expressions into Eq.~(\ref{eq:local_surf}), we find
\begin{equation}
\label{eq:Crslab}
C_z(\boldsymbol{r}) = -\im\,\frac{4\pi}{N_{\boldsymbol{k}}}
\sum_{\boldsymbol{k}} \sum_{vv'c}
  \psi_{v\boldsymbol{k}}(\boldsymbol{r})
  \, X_{vc\boldsymbol{k}} Y^\dagger_{v'c\boldsymbol{k}} \,
  \psi^*_{v'\boldsymbol{k}}(\boldsymbol{r})
\end{equation}
where
\begin{equation}
\label{eq:Xdef}
X_{vc\boldsymbol{k}}
  = \bra{\psi_{v\boldsymbol{k}}}x\ket{{\psi_{c\boldsymbol{k}}}}
  = \frac{\bra{\psi_{v\boldsymbol{k}}}i\hbar v_x\ket{{\psi_{c\boldsymbol{k}}}}}
       {E_{c\boldsymbol{k}}-E_{v\boldsymbol{k}}},
\end{equation}
and similarly for $Y_{vc\boldsymbol{k}}$. The second form in
Eq.~(\ref{eq:Xdef}) makes use of the definition of the velocity operator as
\begin{equation}
\label{eq:v}
\boldsymbol{v}=-\frac{i}{\hbar}\,[\boldsymbol{r},H] \,,
\end{equation}
thereby taming the problematic position operator in the Bloch-state
matrix element.  Averaging Eq.~(\ref{eq:Crslab}) over a
unit cell of area $A$, we find that the contribution of vertical
coordinate $z$ to the Chern number of the slab is
\begin{equation}
\label{eq:czz}
C_z(z)= \frac{-4\pi}{A}\, \im\,\frac{1}{N_{\boldsymbol{k}}}
\sum_{\boldsymbol{k}} \sum_{vv'c}
X_{vc\boldsymbol{k}} Y^\dagger_{v'c\boldsymbol{k}} \,
\rho_{v'v\boldsymbol{k}}(z) \,,
\end{equation}
such that the total slab Chern number is $C_z=\int dz\,C(z)$.
In this equation,
\begin{align}
\rho_{vv'\boldsymbol{k}}(z)
 &= \bra{\psi_{v\boldsymbol{k}}}
    \Big( \int_A d^2r\,\ket{\boldsymbol{r}}\bra{\boldsymbol{r}} \Big)
   \ket{\psi_{v'\boldsymbol{k}}} \nonumber\\
 &= \int_A d^2r\,
      \psi^*_{v\boldsymbol{k}}(\boldsymbol{r})
      \psi_{v'\boldsymbol{k}}(\boldsymbol{r})
\end{align}
is the matrix element in the Bloch representation of the
projector onto the slice through the unit cell at coordinate $z$.

Because we work here in the tight-binding approximation, the sum
over conduction states in Eq.~(\ref{eq:Crslab}) is a finite one
and is easily carried out.  We adopt a diagonal approximation to
the matrix elements of the position operators in the tight-binding
basis, $\bra{j}\boldsymbol{r}\ket{j'}=\bar{\boldsymbol{r}}_j\delta_{jj'}$
(here $j$ labels an orbital located at $\bar{\boldsymbol{r}}_j$
in the unit cell of the slab), in which case the
numerator of Eq.~(\ref{eq:Xdef}) can be evaluated with the help of
\begin{equation}
i\hbar\bra{j}v_x\ket{j'}=(\bar{x}_j-\bar{x}_{j'})H_{jj'}
\end{equation}
(and similarly for $v_y$), again making use of Eq.~(\ref{eq:v}).
Then the contribution of layer $l$ to the Chern number of the
slab follows the form of Eq.~(\ref{eq:czz}), becoming
\begin{equation}
\label{eq:Cz}
C_z(l)= \frac{-4\pi}{A}\, \im\,\frac{1}{N_{\boldsymbol{k}}}
\sum_{\boldsymbol{k}} \sum_{vv'c}
X_{vc\boldsymbol{k}} Y^\dagger_{v'c\boldsymbol{k}} \,
\rho_{v'v\boldsymbol{k}}(l) \,.
\end{equation}
Now
\begin{equation}
\rho_{vv'\boldsymbol{k}}(l) = \sum_{j\in l}
      \psi^*_{v\boldsymbol{k}}(j) \psi_{v'\boldsymbol{k}}(j)
\end{equation}
is the Bloch representation of the projection onto layer $l$,
where the sum is over orbitals $j$ belonging to that layer.
All of the ingredients needed to compute Eq.~(\ref{eq:Cz}) are
thus easily evaluated, and the contribution of that layer to the
AHC is just $\sigma(l)=(e^2/h)C_z(l)$.

As explained in Ref.~[\onlinecite{rauch-prb18}], the object
$F_{vv'\boldsymbol{k}}=(XY^\dagger)_{vv'\boldsymbol{k}}$ is the
covariant metric-curvature tensor.  Because $\rho_{\boldsymbol{k}}(l)$
is a Hermitian matrix, the imaginary part operation in Eq.~(\ref{eq:Cz})
filters out only the antihermitian part of $F_{\boldsymbol{k}}$,
which is $-2i$ times the covariant Berry curvature tensor
$\Omega_{vv'\boldsymbol{k}}$. Thus, formally we have that
\begin{equation}
C_z(l)= \frac{2\pi}{A}\frac{1}{N_{\boldsymbol{k}}}\sum_{\boldsymbol{k}}
    {\rm Tr}[\Omega_{\boldsymbol{k}}\rho_{\boldsymbol{k}}(l)]
\end{equation}
or, when converted to a Brillouin zone integral,
\begin{equation}
C_z(l)= \frac{1}{2\pi}  \int_{\rm BZ} d^2k\,
    {\rm Tr}[\Omega_{\boldsymbol{k}}\rho_{\boldsymbol{k}}(l)] \,.
\end{equation}
While this form is not as convenient for computational purposes as
Eq.~(\ref{eq:Cz}), it is more intuitive.  For example, since
$\sum_l \rho_{vv'\boldsymbol{k}}=\delta_{vv'}$, this immediately leads to
$C_z=\sum_l C_z(l)=(1/2\pi)\int_{\rm BZ} {\rm Tr}[\Omega_{\boldsymbol{k}}]$
as it should.

In summary, we use Eq.~(\ref{eq:Cz}) to compute the contribution $C_l(z)$ of each surface layer to the surface AHC.  If these vanish as $z$ goes into the interior, then we simply sum the surface-layer contributions and multiply by $e^2/h$ to get the surface AHC. If instead the $C_l(z)$ oscillate in the bulk, then a coarse-graining procedure is used to isolate the surface contribution, as will be described in Sec.~\ref{sec:sahc}.

\section{Minimal model on the pyrochlore lattice}
\label{sec:pyro}
In this section we motivate and introduce a class of tight-binding models of spinors on the pyrochlore lattice.  

\subsection{The pyrochlore lattice: A playground for topological phenomena }

\begin{figure}
\centering\includegraphics[width=\columnwidth]{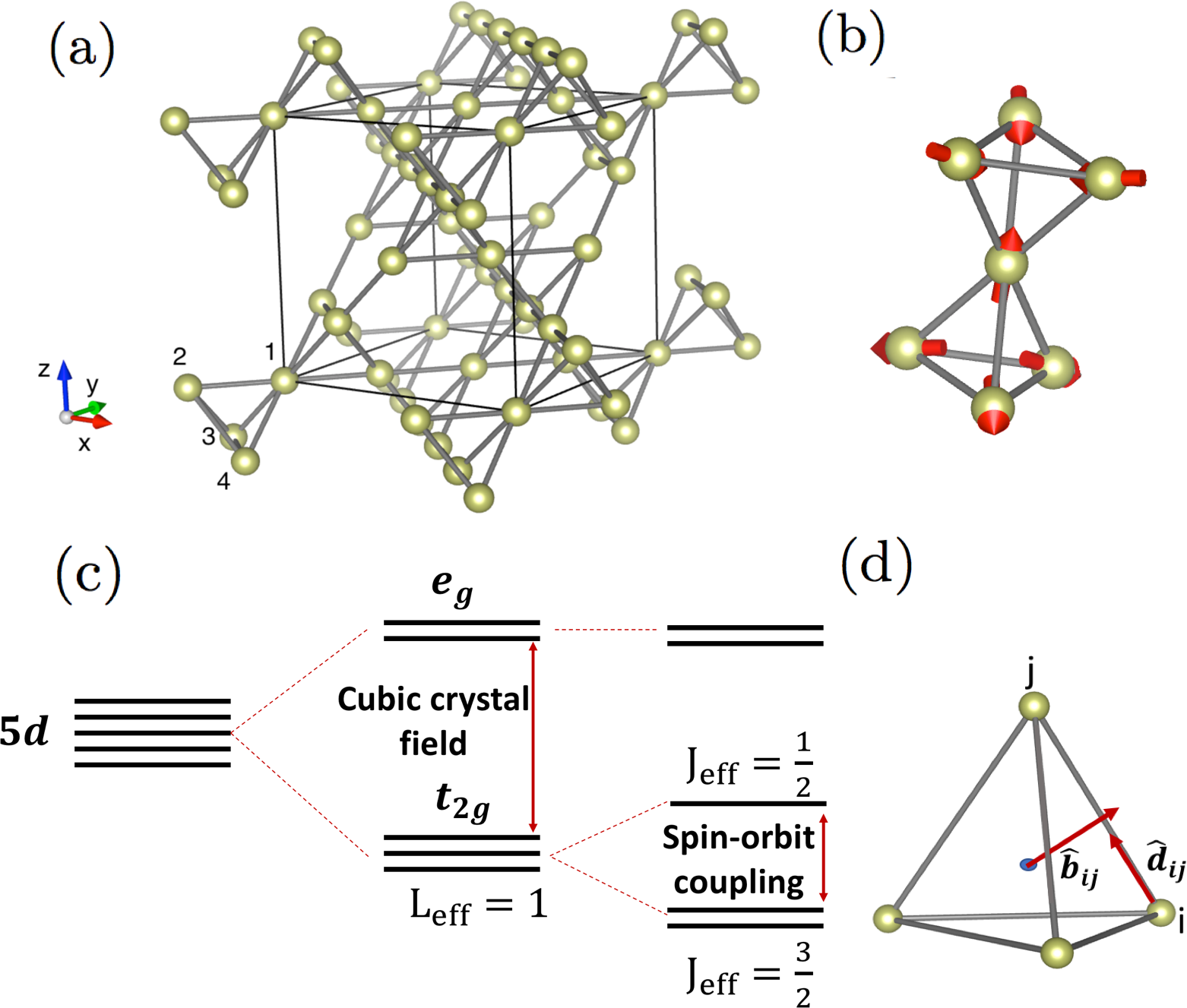}
\caption{(a) The pyrochlore lattice, comprised of an fcc lattice with a four-point basis (denoted by the numbers 1-4), forming a corner sharing tetrahedral network. (b)  All-in-all-out configuration predicted for pyrochlore iridates. (c) The Ir 5$d$ level splitting in pyrochlore iridates due to the crystal field and the spin-orbit coupling. (d) Graphical definition of $\boldsymbol{\hat{b}}_{ij}$ and $\boldsymbol{\hat{d}}_{ij}$}
\label{fig:tetra}
\end{figure}

In both the $A_2B_2$O$_7$ pyrochlore and $AB_2$O$_4$ spinel crystal structures, the $B$-site atoms form a simple inversion-symmetric pyrochlore lattice as shown in Fig.~\ref{fig:tetra}(a).  Here we are interested in systems in which the metal atom on the $B$ is magnetic, while the one on the $A$ site is not.  This means that our considerations will be limited to materials in which the $A$-site atom does not have a partially filled $f$ shell, or if it does, we are above the $f$-moment ordering temperature. In the case that a 5$d$ transition metal occupies the $B$ site, for example in pyrochlore $A_2$Ir$_2$O$_7$ iridates or spinel $A$Os$_2$O$_4$ osmates, the 5$d$ electrons will exhibit strong spin-orbit coupling that can lead to band inversions. In fact, these 5$d$ electron systems have drawn a lot of attention recently, since the interplay between electronic correlation and that of spin-orbit coupling leads to a variety of topological and magnetic phases.~\cite{pesin-n10,guo-prl09,wan-prl12,wan-prb11,krempa-prb12,chen-prb12,yang-prb10,kurita-psj11, yamaji-prx14, yamaji-prb16, tomiyasu-psj12,sagayama-prb13,disseler-prb14,donnerer-prl16,zhang-prl17}

In the case of osmium spinels, it was shown~\cite{wan-prl12} that for a reasonable range of the on-site Coulomb correlation $U$, the ground state can be a FM axion insulator. On the other hand, in the case of pyrochlore iridates it was shown~\cite{krempa-prb12,chen-prb12,wan-prb11} that electronic interactions can lead to topological phases with a noncollinear AIAO spin ordering, as shown in Fig.~\ref{fig:tetra}(b). Even though various experiments~\cite{tomiyasu-psj12,sagayama-prb13,disseler-prb14,donnerer-prl16} have confirmed that the ground state exhibits the AIAO phase, it was recently predicted~\cite{zhang-prl17} that pyrochlore iridates are topologically trivial in this phase.

\subsection{The model} \label{III.B.}

We can arrive at a simple tight-binding description of the pyrochlore iridates by considering the electronic states close to the Fermi energy.  The oxygen octahedra surround the Ir$^{4+}$ ions, creating a strong cubic crystal field that splits the Ir 5$d$ orbitals into $t_{2g}$ and $e_g$ multiplets. The effective angular momentum of the $t_{2g}$ levels is $L_{\textrm{eff}}=1$, so when the onsite SOC is ``turned on," it splits the $t_{2g}$ multiplet into an effective pseudospin $J_{\textrm{eff}} = 1/2$ doublet and a corresponding $J_{\textrm{eff}} = 3/2$ quadruplet,\citep{kim-prl08} as illustrated in Fig.~\ref{fig:tetra}(c). Since Ir$^{4+}$ has five electrons in the 5$d$ shell, the $J_{\textrm{eff}} = 3/2$ quadruplet is filled, while the $J_{\textrm{eff}} = 1/2$ doublet is half-filled. Finally, the Fermi energy lies sufficiently far from the $J_{\textrm{eff}} = 3/2$ level so that, for the purposes of a minimal model, we can safely ignore it and describe the system in terms of a single spinor degree of freedom per site.

In the following we consider only nearest-neighbor interactions, with
\begin{equation} \label{eq:H_t}
 H_t = t \sum_{\langle ij \rangle \sigma}  c^\dagger_{i\sigma}c_{j\sigma} + \text{h.c.}   
\end{equation}
describing the spin-independent hoppings. For the spin-dependent interactions, we begin by imposing time-reversal symmetry, in which case the spin-orbit induced hopping between nearest neighbors takes the form
\begin{equation} \label{eq:H_l}
 H_{\lambda} =  \lambda \sum_{ \langle ij \rangle \alpha \beta} i \sqrt{2} c^\dagger_{i\alpha} \, \boldsymbol{\hat{b}}_{ij} \times \boldsymbol{\hat{d}}_{ij}\cdot \boldsymbol{\sigma_{\alpha\beta}} \, c_{j\beta}+ \text{h.c.}
\end{equation}
Here $\boldsymbol{\sigma}=(\sigma_x,\sigma_y,\sigma_z)$ are the Pauli spin matrices, $\boldsymbol{\hat{d}}_{ij}$ is the unit vector connecting site $i$ to site $j$, and $\boldsymbol{\hat{b}}_{ij}$ is the unit vector from the center of a tetrahedron to the midpoint of the bond $\langle ij \rangle$ (see Fig.~\ref{fig:tetra}(d) for a graphical definition).\cite{kurita-psj11,guo-prl09} This structure is imposed by the presence of a mirror symmetry across the plane containing vectors $\boldsymbol{\hat{d}}_{ij}$ and $\boldsymbol{\hat{b}}_{ij}$, which ensures that only the component of $\boldsymbol{\sigma}$ normal to this mirror plane can appear. The factor of $i$ in front is required by time-reversal symmetry, which reverses the signs of the Pauli matrices in addition to applying complex conjugation.  

Next, in order to break time-reversal symmetry and model the magnetic ordering in these systems, we consider an onsite Zeeman term
\begin{equation} \label{eq:H_zeeman}
 H_\Delta = \Delta \sum_{i} \boldsymbol{\hat{n}}_i \cdot \boldsymbol{\sigma}c^\dagger_{i}c_{i} ,
\end{equation}
where $\boldsymbol{\hat{n}}_i$, $i=\{1,2,3,4\}$ are the vectors on each site describing the spin configuration.%
\footnote{Broken time-reversal symmetry also allows an $\hat{n}_{ij}\cdot\boldsymbol{\sigma}c^\dagger_{i}c_{j}$ term to appear in the nearest-neighbor hopping, but the modification of the on-site Zeeman term is much more directly motivated.}
As mentioned earlier, the appearance of magnetic moments is a consequence of on-site Hubbard repulsion in the many-body Hamiltonian,\cite{krempa-prb12,wan-prb11} and the AIAO long-range order is established by the pattern of intersite exchange interactions.\cite{chen-prb12}  Here we assume the presence of the AIAO magnetic order and regard our Hamiltonian in Eq.~(\ref{eq:H_zeeman}) as resulting from a mean-field approximation to the correlated system of interest.

\section{Results}
\label{sec:results}
%
In the following, we first consider in detail the case of AIAO spin ordering, and then briefly describe how the results change when considering the FM ordering as a function of orientation of the magnetization.
The calculation of bulk energy bands, surface states and Wannier bands was done using the built-in tools provided by the open-source code package PythTB.\cite{pythtb} \Changes{For the calculation of the
layer-resolved anomalous Hall conductivity, we implement an
extension package of PythTB that has been contributed to
an open-source repository.\footnote{\Changes{PythTB scripts for computing
   layer-resolved contributions to the anomalous Hall conductivity
   are available at \href{http://www.physics.rutgers.edu/~dhv/pythtb-resources/}{http://www.physics.rutgers.edu/$\sim$dhv/pythtb-resources}
   }}}

\subsection{All-in-all-out spin configuration}
\label{sec:aiao}

\subsubsection{Bulk energy bands}
\begin{figure}
\centering\includegraphics[width=\columnwidth]{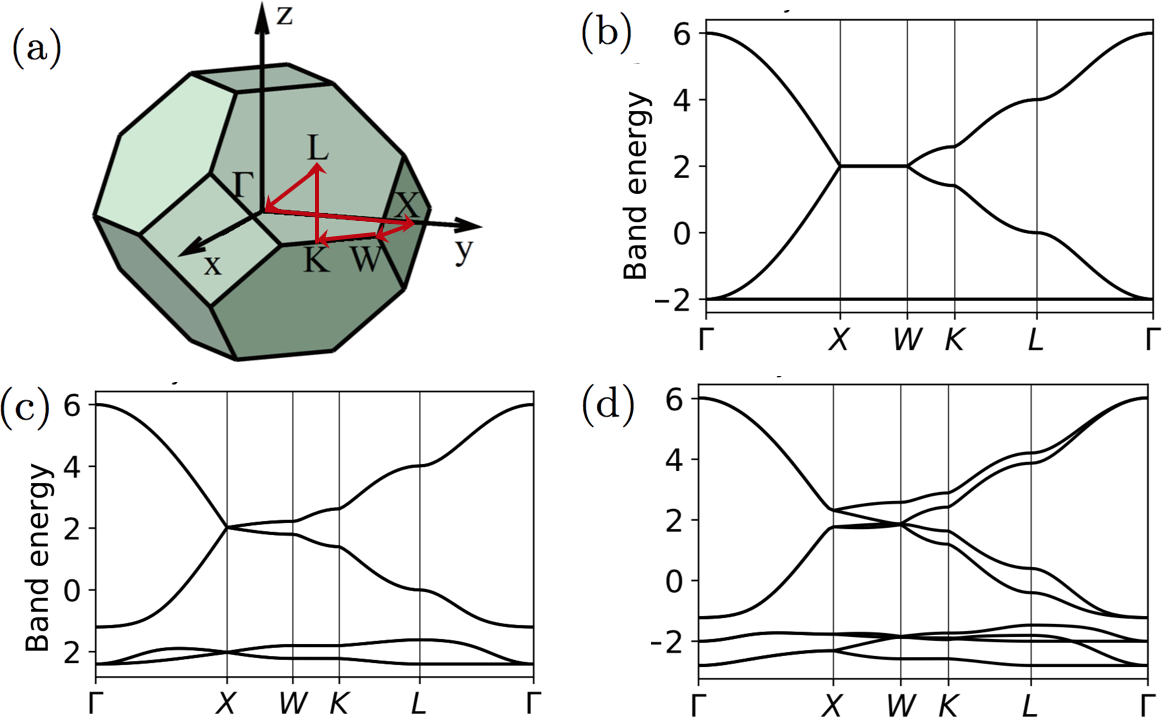}
\caption{(a) The first Brillouin zone with the path along which the bulk energy bands are calculated. (b)-(d) Energy bands for different values of the parameters $(t,\lambda,\Delta)$. (b) $(1,0,0)$. (c) $(1,0.1,0)$. (d) $(1,0.1,0.2)$.}\label{fig:bandstructure}
\end{figure}

As we have seen in section \ref{III.B.}, the Hamiltonian consists of three terms $H(t,\lambda,\Delta) = H_t+H_{\lambda}+H_{\Delta}$. To get a better picture of the nature of the interactions, we turn our attention to the energy bands.  First, we turn on only the spin-independent hoppings, $H=H(t,0,0)$, obtaining the results shown in Fig.~\ref{fig:bandstructure}(b).  Since spin up and down are equivalent, all bands are doubly degenerate, and the Hamiltonian is easily diagonalized to obtain the four solutions
\begin{equation} \label{eq:Ht_energies}
\begin{split}
 E^{(1,2)}_{\boldsymbol{k}} & = -2t \\
 E^{(3,4)}_{\boldsymbol{k}} & = 2t[1 \pm \sqrt{1+A_{\boldsymbol{k}}} ],
 \end{split}
\end{equation}
where
\begin{equation} \label{eq:Ak}
A_{\boldsymbol{k}} = \cos{2k_x}\cos{2k_y}+\cos{2k_y}\cos{2k_z}+\cos{2k_z}\cos{2k_x}.
\end{equation}
(The extra degeneracy of the first two bands is a well-known artifact of this minimal model.) Since we are at half filling, this Hamiltonian describes a metal with a quadratic band touching at $\Gamma$.

Next we turn on the $\T$-invariant intersite spin-orbit coupling term, so that now $H=H(t,\lambda,0)$. Even though the Hamiltonian contains spin-dependent interactions, the combination of $\T$ and $\I$ symmetries forces the bands to remain doubly degenerate everywhere.  However, the other degeneracies are lifted except at some high-symmetry points, as shown in Fig.~\ref{fig:bandstructure}(c) for $\lambda=0.1t$, where it is also clear that a global gap has opened between the valence and conduction bands.  This insulating state cannot be connected adiabatically to the atomic limit and corresponds to a strong TI~\cite{kurita-psj11,guo-prl09} with $\theta=\pi$.

Finally, we break $\mathcal{T}$ by turning on $\Delta$.  For $\Delta$ small enough that the gap does not close, as in Fig.~\ref{fig:bandstructure}(d), $\theta$ must remain equal to $\pi$. Now the topological phase is protected by inversion, and the system is an axion insulator. The breaking of $\mathcal{T}$ means that spin up and down electrons no longer disperse in the same way, and all eight bands are nondegenerate except at some symmetry points or lines in the Brillouin zone.

\subsubsection{Phase diagram} 
\label{sec:phasediagram}
\begin{figure}
\centering\includegraphics[width=0.9\columnwidth]{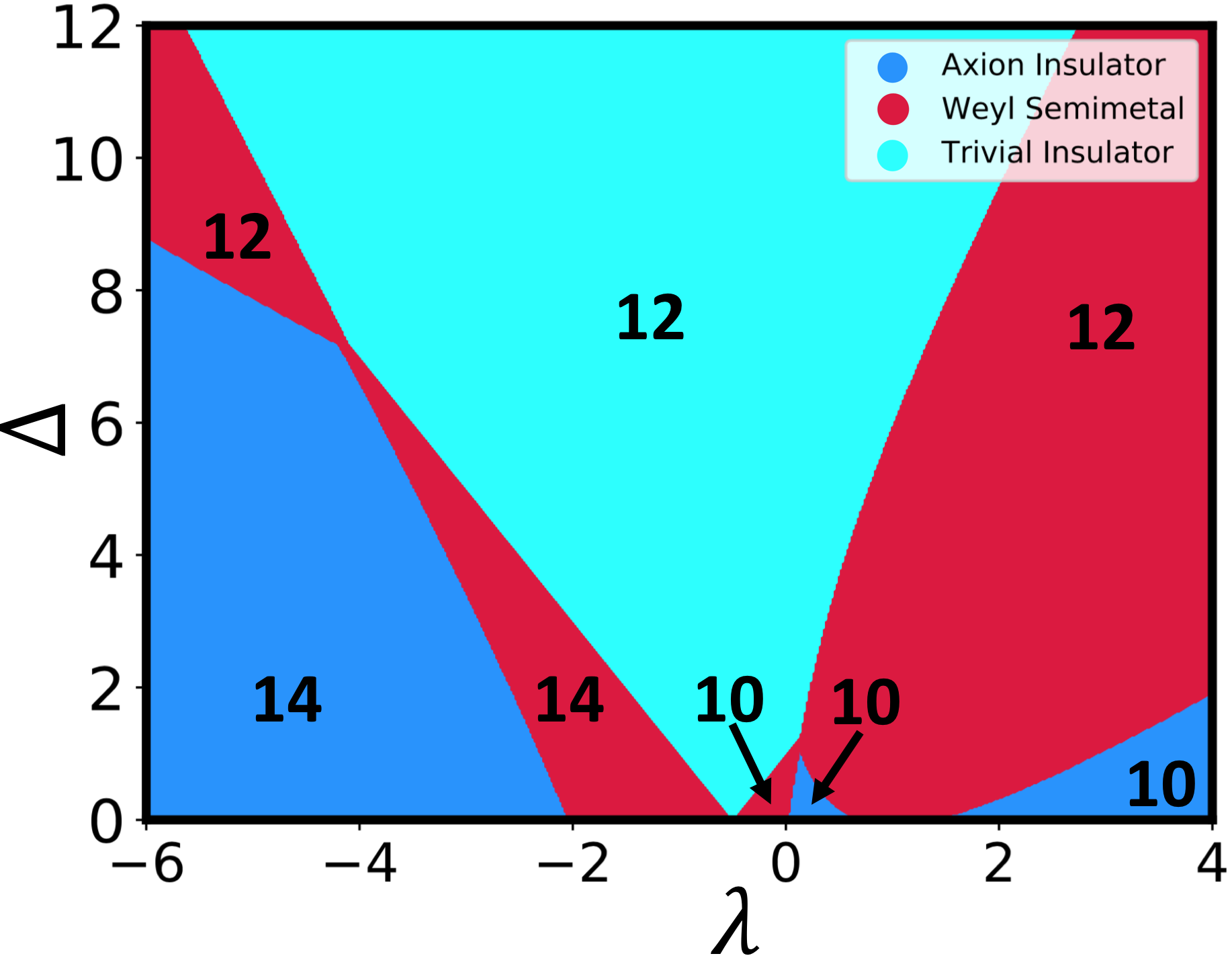}
\caption{Phase Diagram for $t=1.0$ with the distribution of the total number of odd parity states at the TRIM. The Hamiltonian remains invariant under $\Delta \rightarrow -\Delta$ so it is sufficient to consider positive $\Delta$. On the other hand, $H(-t,-\lambda,\Delta) = -H(t,\lambda,\Delta)$ so for $t=-1.0$ the occupied and unoccupied bands are interchanged. The phase diagram is nevertheless the same, since we assume half filling and the 8 bands constitute a trivial insulator.}
\label{fig:phase_diagram}
\end{figure}

Even though the expression for $\theta$ in Eq.~(\ref{eq:axioncoup}) is elegantly written in its integral form, applying it is computationally expensive and somewhat problematic.\citep{coh-prb13} Fortunately, things are greatly simplified if one has a knowledge of the parities at the time-reversal invariant momenta (TRIM). As shown by Fu and Kane,\cite{fu-prb07} a counting of the parities at the TRIM determines the $\mathbb{Z}_2$ invariant when $\T$ and $\I$ are both present. Later it was shown that either time reversal~\cite{qi-prb08} or inversion~\citep{essin-prl09+e} alone can quantize the axion coupling $\theta$, and the parity-counting rule was generalized to the inversion-only case.\cite{hughes-prb11, turner-prb12} This rule states that if the material is insulating and is not a Chern insulator,%
\footnote{In general, a three-dimensional insulator is characterized by three integer Chern indices; if any of these are nonzero, the system is a Chern insulator, and it displays a quantized anomalous Hall conductivity.  We never encountered such a phase in our model system. Indeed, the axion phases in our model are adiabatically connected to a strong TI phase, for which the Chern indices vanish by $\T$ symmetry.} then $\theta = \pi$ if and only if the total number of odd-parity states (NOPS) at the TRIM is twice an odd integer.

Using this rule, we iterate over the parameters of our model to acquire the phase diagram shown in Fig.~\ref{fig:phase_diagram}, where the numbers indicate the NOPS. Here we use the topological definition of an insulator, insisting only on a global direct gap between the four valence and four conduction bands.  The trivial and non-trivial insulating states are separated by ``Weyl Semimetal'' regions that are represented by the red color.

The boundaries in Fig.~\ref{fig:phase_diagram} correspond to crossings of eigenvalues, i.e., band inversions, at the $\Gamma$ and $L$ points. All states at $\Gamma$ have the same (even) parity, and the parity counts at the $X$ points never change, so the NOPS is determined by the parity sum over the four $L$ points.  Thus, the boundaries with a discontinuity in the NOPS involve band inversion at $L$, while the others are at $\Gamma$, and the Weyl semimetal regions lie in between.

Our picture, then, is as follows. Choose any path that starts from the axion phase, enters the Weyl semimetal phase, and ends in the trivial phase in Fig.~\ref{fig:phase_diagram}. If you enter the Weyl region without changing the total NOPS, i.e., by passing through a Weyl semimetal region with 10 or 14 total NOPS, then you find that eight Weyl points are created at $\Gamma$ when crossing the boundary. These then separate and migrate along the eight equivalent $\Gamma \rightarrow L$ lines, and finally meet and annihilate in pairs at the four $L$ points at the crossing into the trivial insulator phase. If the path passes through a Weyl region with 12 total NOPS, on the other hand, the ordering of events is reversed. 
\subsubsection{Surface states}

\begin{figure}
\centering\includegraphics[width=\columnwidth]{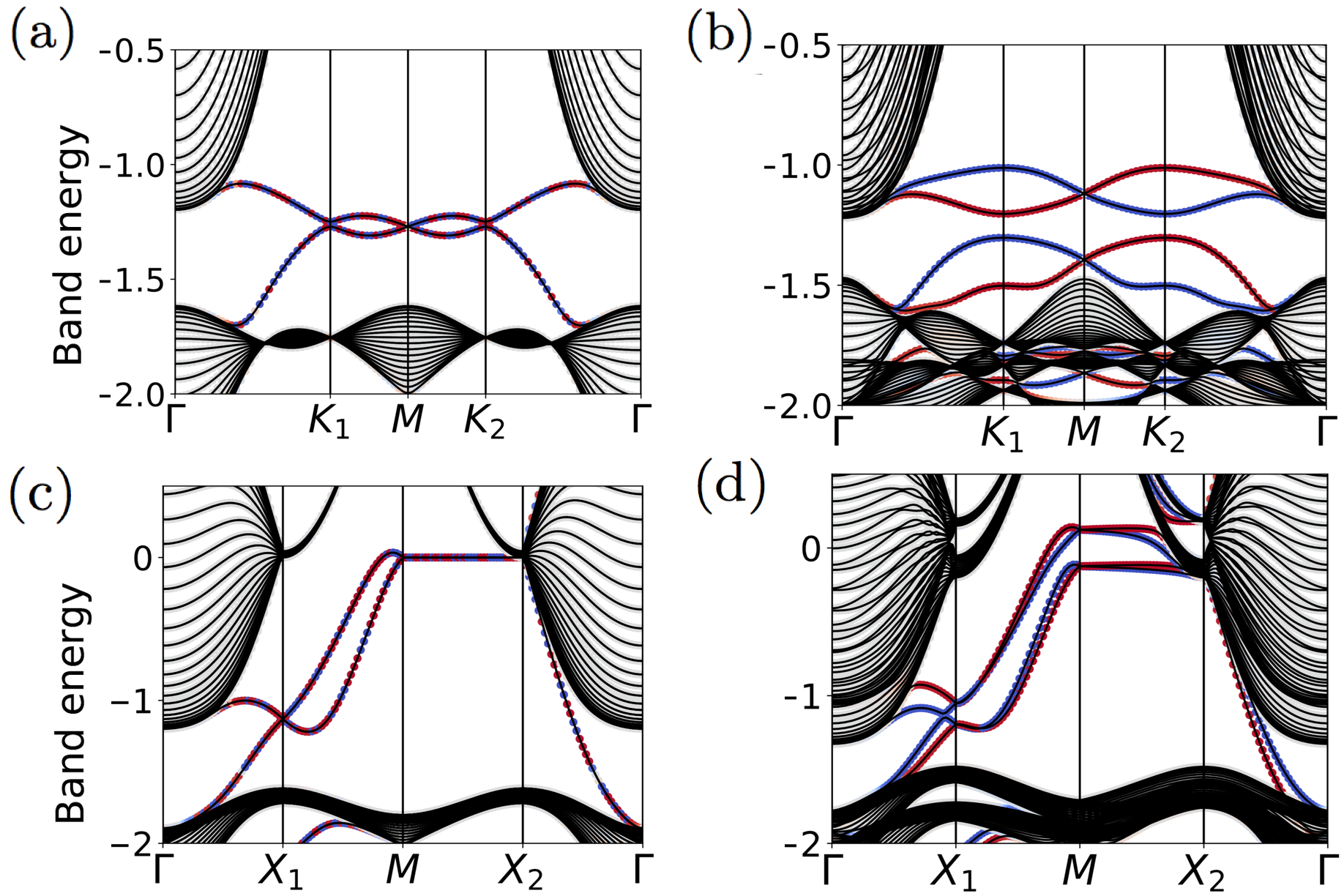}
\caption{Surface band structures for (a-b) $(111)$ surfaces, and (c-d)   $(001)$ surfaces. Blue (red) bands are localized on the bottom (top) surface. (a,c) Strong TIs, $H(t,\lambda,\Delta)=(1,0.1,0)$, with protected metallic surface states. (b,d) Axion insulators, $H(t,\lambda,\Delta)=(1,0.1,0.4)$; note gapped surface states on the (111) surface.}
\label{fig:surface_states}
\end{figure}

We create slabs that are in the axion-insulator phase and then use Eq.~(\ref{eq:Cz}) to find the magnitude and sign of the surface AHC. Since, the bulk AHC is zero, and the surface AHC is not defined except for the component normal to the surface, we simply refer to it as $\sigma_{\text{AHC}}$. For the surface AHC to be half-quantized, however, the surface band structure has to be gapped. In the axion phase there is no topological protection forcing surface states to cross the gap, but neither is an open gap guaranteed.

A slab along the $(111)$ direction consists of alternating kagome and triangular atomic layers, so the 2D Brillouin zone is hexagonal. Fig.~\ref{fig:surface_states}(a) shows the protected metallic surface band structure of a strong TI for a slab that terminates on a triangular layer. When the $\mathcal{T}$-breaking $\Delta$ term is turned on, the surface states become gapped as can be seen in Fig.~\ref{fig:surface_states}(b). 

Along $(001)$ the slab consists of tetragonal layers rotated $90^{\circ}$ with respect to each other, so the 2D Brillouin zone is tetragonal. As before, we choose the top and bottom layers to have the same orientation, so that the slab as a whole has inversion symmetry. In contrast with the $(111)$ case, the surface states do not become gapped in the axion phase, as shown in Fig.~\ref{fig:surface_states}(d). However, we have found that we can obtain an insulating surface by artificially increasing the strength of the Zeeman term on the surface atoms only, and this modified Hamiltonian will be used for some of the results presented below.

\subsubsection{Surface anomalous Hall conductivity}
\label{sec:sahc}

\begin{figure}
\centering\includegraphics[width=\columnwidth]{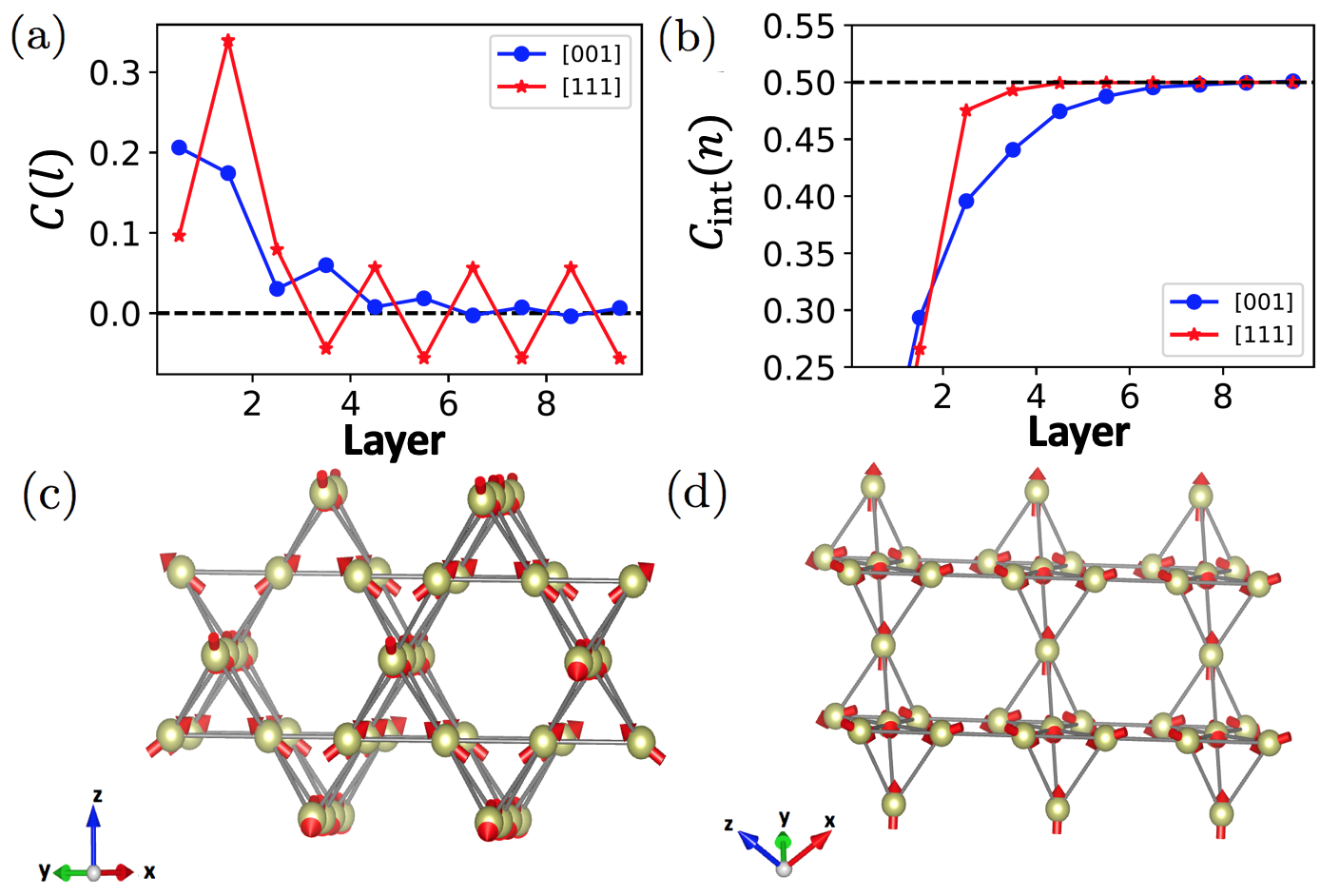}
\caption{(a) Layer-resolved partial Chern number $C(l)$ of \equ{Cz} as a function of layer depth $l$ for slabs along $(111)$ and $(001)$ directions. For both slabs the bulk parameters are $(t,\lambda,\Delta) = (1.0,0.1,0.4)$ corresponding to the axion phase. In the case of the $(001)$ direction the surface Zeeman term is modified to $\Delta_{\textrm{surf}} = 0.8$ in order to gap the surface states. (b) Integral of  $C(l)$ over a surface region extending to depth $n$ as computed from \equ{Cn}. (c-d) Sketches of (001) and (111) slab orientations respectively for the AIAO spin configuration.}
\label{fig:surfaceahc}
\end{figure}

We apply Eq.~(\ref{eq:Cz}) for the slabs constructed along $(111)$ and $(001)$ directions and find the partial Chern numbers $C_z(l)$ of each layer, which we now denote just as $C(l)$ for brevity.  The results are shown in  Fig.~\ref{fig:surfaceahc}(a) for slabs consisting of 20 layers and using a \Changes{40x40 k-mesh sampling}. In both cases, the contribution to the AHC comes from the first few layers, and as one goes deeper in the bulk, the partial Chern number oscillates around zero. Note that the kagome layers contain three times as many atoms as the triangular layers Fig.~\ref{fig:surfaceahc}(c), which may explain why the second layer along the $(111)$ direction has approximately three times the partial Chern number of the first layer.  In Fig.~\ref{fig:surfaceahc}(b) we integrate the partial Chern number and show that $\sigma_{\text{AHC}}$ is quantized to half the quantum of conductance. To tame the oscillatory behavior of the bulk layers we carry out the sum to a depth $n$ according to
\Changes{
\begin{equation}
\label{eq:Cn}
C_{\textrm{int}}(n) = \sum_{l=0}^{n-1}\bar{C}_{l,l+1}
\end{equation}
where $\bar{C}_{l,l+1}$ is the coarse-grained average of two neighboring layers: $\bar{C}_{l,l+1}=[C(l)+C(l+1)]/2$ with the exception $\bar{C}_{0,1}=C(1)/2$.  In this way each layer is counted once except for the layer at depth $n$, which is counted with weight 1/2.  This can be regarded as an application of the sliding window averaging method as described, e.g., in Appendix~C of Ref.~\onlinecite{Resta2018}.  Since these pairwise layer averages decay to zero in the bulk, the sum converges rapidly with depth $n$.}

In both slabs we see that each time we add or remove a layer, the sign of $\sigma_{\text{AHC}}$ flips. For the (001) slab, we can understand this as follows. Adding a layer is equivalent to a $90^\circ$ rotation around $[001]$ and a half-lattice-constant translation along $[001]$ followed by $\T$, as is evident from Fig.~\ref{fig:surfaceahc}(c). Neither the rotation nor the translation affects $\sigma_{\text{AHC}}$, but $\T$ will flip its sign. Now for the (111) slab, things are less obvious since triangular and kagome terminations are inequivalent. Nevertheless,  we can heuristically understand the sign flip by the expectation that the sign of the surface AHC tracks the sign of the surface magnetization, which reverses every time we add or remove a layer [see Fig.~\ref{fig:surfaceahc}(c)].

\subsubsection{Wannier bands}
\label{sec:wannier_bands}
\begin{figure}
\centering\includegraphics[width=\columnwidth]{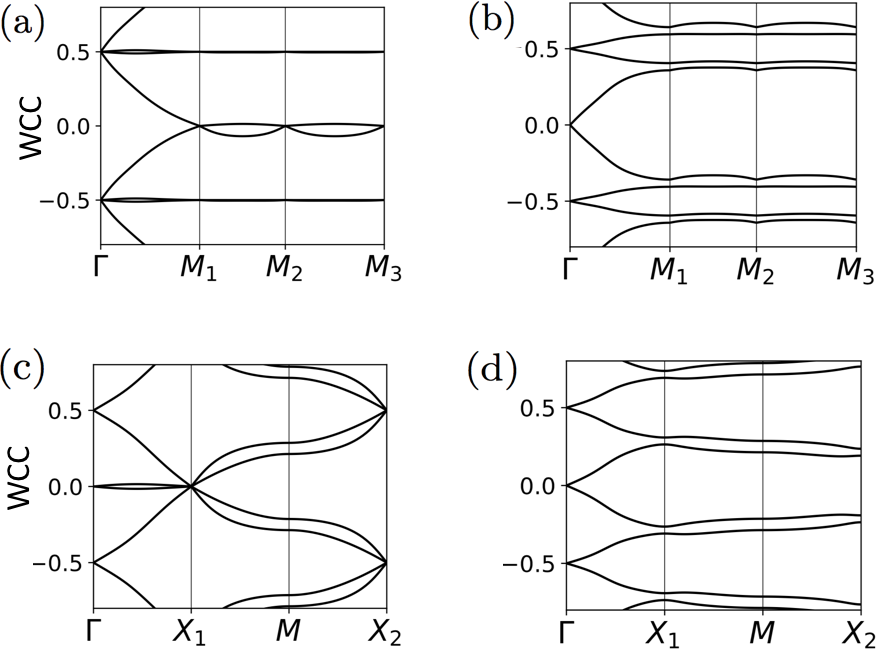}
\caption{WCCs for an axion insulator along the 4 projected TRIM for Wannierization directions (111) (a),(b) and (001) (c),(d). Figures (a),(c) correspond to a region in the phase diagram with 10 odd parity states at the TRIMs, where (b),(d) correspond to 14 odd parity states.}
\label{fig:wannier_bands}
\end{figure}

An analysis based on Wannier functions (WFs) offers an alternative description of the ground state of a periodic crystal. Namely, one carries out a unitary transformation from the Bloch functions $\ket{\psi_{n\mathbf{k}}} =e^{i\mathbf{k}\cdot\mathbf{r}}\ket{u_{n\mathbf{k}}}$ to a set of localized WFs
\begin{equation} \label{eq:WFs}
\ket{W_n(\mathbf{R})} = \frac{1}{(2\pi)^3}\int_{BZ} d\mathbf{k} \, e^{-i\mathbf{k}\cdot\mathbf{R}} \ket{\psi_{n\mathbf{k}}}
\end{equation}
labeled by a cell index $\mathbf{R}$ and a band-like index n, such that WFs at different $\mathbf{R}$ are translational images of one another. More generally one adopts the freedom to apply an arbitrary $\mathbf{k}$-dependent unitary transformation to the occupied Bloch states at each $\mathbf{k}$, $\ket{\tilde{\psi}_{n\mathbf{k}}} = \sum_m U_{mn}(\mathbf{k})\ket{\psi_{m\mathbf{k}}}$ before applying \equ{WFs}, so the WFs are strongly non-unique.

In one dimension, however, there is a unique gauge that minimizes the spread functional of the WFs.\citep{marzari-prb97, marzari-rmp12} Following Refs.~\onlinecite{soluyanov-prb11,yu-prb11,taherinejad-prb14}, we take advantage of this fact by constructing hybrid WFs that are Wannier-like in one of the dimensions and Bloch-like in the remaining ones. For example in 3D these take the form
\begin{equation} \label{HWFs}
\ket{W_{nR_z}(k_x,k_y)} = \frac{1}{2\pi}\int^{\pi}_{-\pi} dk_z \, e^{-ik_zR_z} \ket{\psi_{n\mathbf{k}}}
\end{equation}
where we have chosen $\mathbf{\hat{z}}$ as the Wannierization direction. One can plot the Wannier charge centers (WCC) in the home unit cell $R_z=0$
\begin{equation} \label{WCC}
\bar{z}_n(k_x,k_y) = \bra{W_{n0}}\hat{z}\ket{W_{n0}}
\end{equation}
and examine how the WCCs ``flow", in the sense of connectedness along the Wannierization direction.

This kind of analysis of the flow of the hybrid WCCs, otherwise known as Wilson loop eigenvalues, has been used extensively to study the topological properties of crystals.\citep{taherinejad-prl15,alexandradinata-prb14,taherinejad-prb14,olsen-prb17,benalcazar-prb17} For example, Taherinejad \textit{et al.}~\citep{taherinejad-prb14} demonstrated how the Wannier bands of a strong TI must exhibit flow, no matter which direction is chosen for the Wannierization.

One may wonder, is the same is true of an axion insulator?  Both axion insulators and strong TIs have the same non-trivial axion index, but axion insulators have broken $\mathcal{T}$ symmetry. It turns out that the answer is no: a non-trivial axion index does not require flow.

To demonstrate this, in Fig.~\ref{fig:wannier_bands} we plot the WCCs along paths connecting the four projected TRIM. We do this for two different Wannierization directions and two different parameter sets that describe an axion insulator. Figs.~\ref{fig:wannier_bands}(b-c) show axion insulators that do not exhibit Wannier flow. Whether there is flow or not depends on how the number of odd parity states is distributed at the four projected TRIM. The nature of the Wannier bands in centrosymmetric crystals, and a discussion of how the axion $\zt$ invariant can often be deduced by an inspection of this Wannier structure, will be presented in a forthcoming publication.
\subsection{Ferromagnetic spin configuration}
\label{sec:fm}
\begin{figure}
\centering\includegraphics[width=\columnwidth]{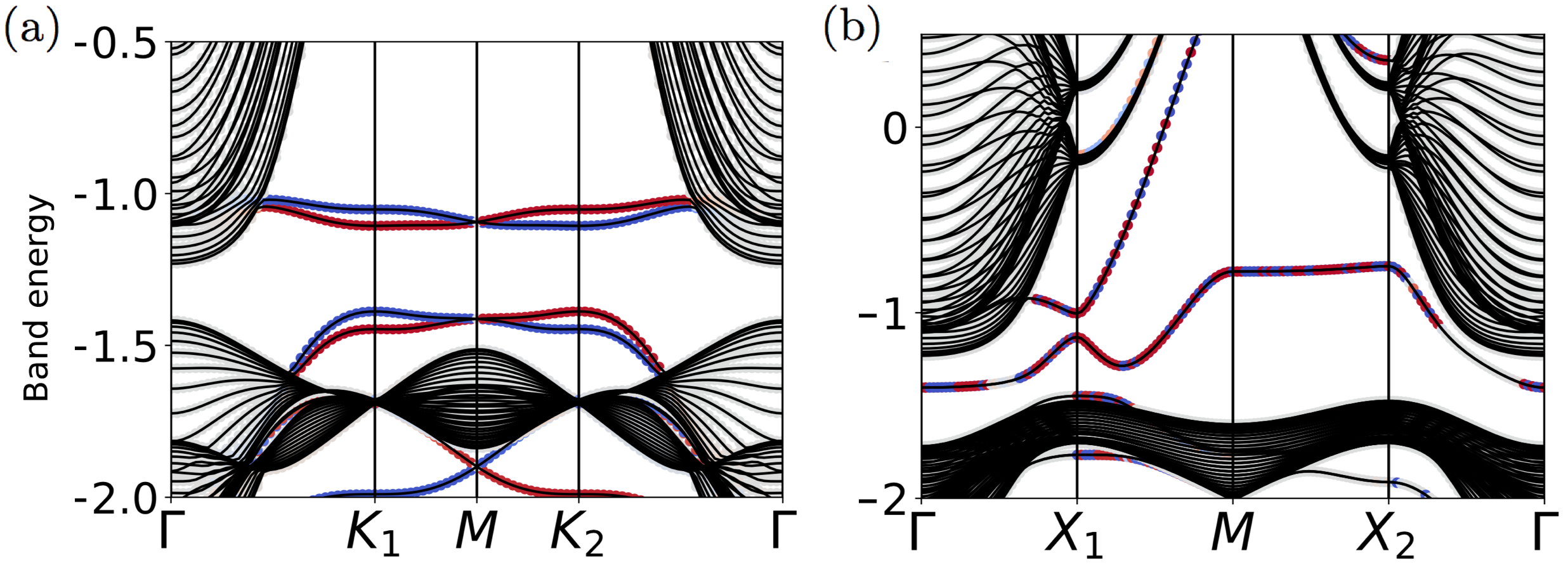}
\caption{Gapped surface states in the FM axion-insulator phase. Slabs along (a) (111) (b) (001) directions. In both cases the magnetization is perpendicular to the surface.}
\label{fig:ferro_states}
\end{figure}

It has been theoretically argued~\cite{wan-prl12} that the ground state of osmium spinels, which share the same crystal structure as pyrochlores, is an FM axion insulator. From a topological point of view the FM and the AIAO spin configurations behave in a similar way, since both respect inversion symmetry. States that were in the strong TI phase for $\Delta = 0$ will become axion insulators when the ferromagnetism is turned on, up to a critical Zeeman field of $\Delta_{c_1}$ at which the bulk gap closes. Eventually, for $\Delta \gg t$ and $\lambda$, the system is a strongly spin-polarized insulator, with a gap separating spin-up from spin-down states, so that a second critical field $\Delta_{\textrm{c}2} > \Delta_{\textrm{c1}}$ must exist where the gap reopens. As in the AIAO case, we find that the surface states can be gapped in the FM axion-insulator phase, at least when the magnetization is normal to the surface, as shown in Fig.~\ref{fig:ferro_states}.

\section{Surface Phenomena}
\label{sec:surfacephenomena}

In the previous sections we have constructed a model for an axion insulator and introduced a computational tool that enables us to calculate the surface AHC. In a sense, then, we have in hand a kind of virtual laboratory that we can use to explore the various phenomena that can occur on the surface of an axion insulator.

\subsection{Surface AHC and magnetization direction}
\label{sec:fmmag}

\begin{figure}
\centering\includegraphics[width=\columnwidth]{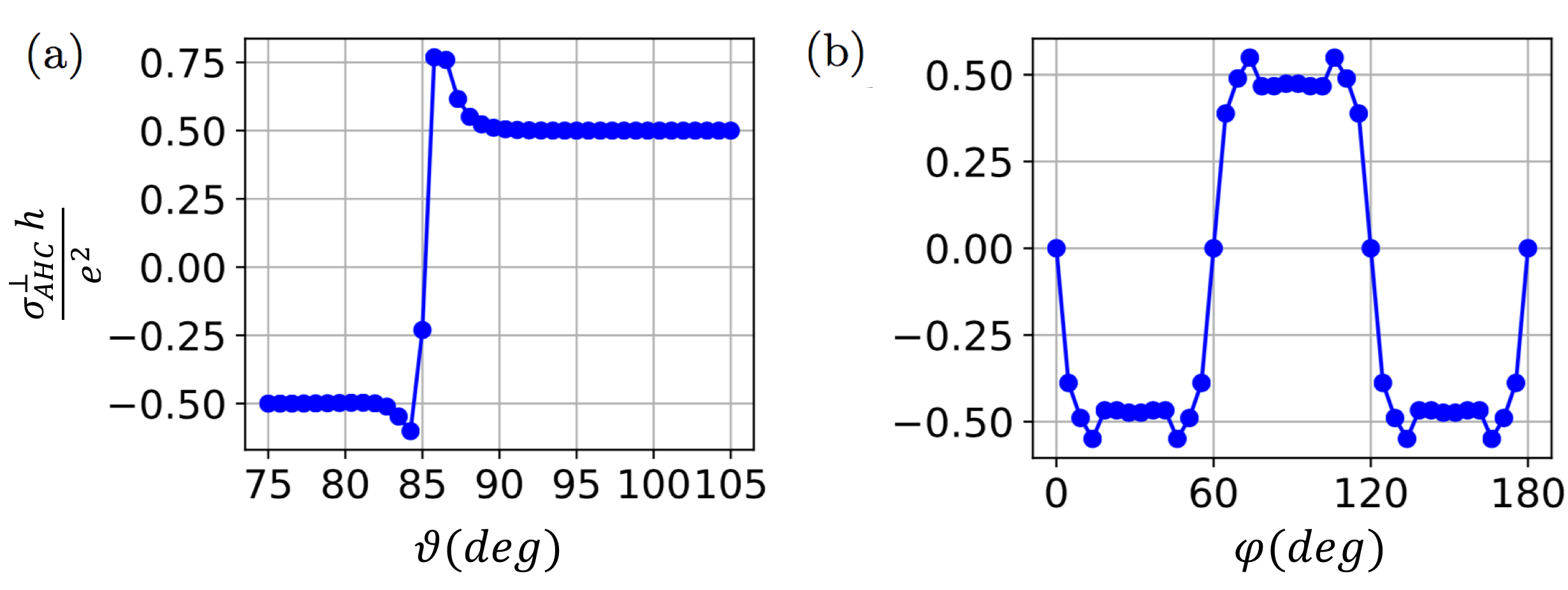}
\caption{Surface AHC for a $(111)$ slab as the magnetization is rotated: (a) in a plane perpendicular to the surface; (b) in the plane of the surface. The polar angle $\vartheta \!=\!0$ corresponds to the $[\bar{1}\bar{1}\bar{1}]$ direction while the azimuthal angle $\varphi \!=\!0$ corresponds to the $[01\bar{1}]$ direction ($[01\bar{1}]$ is perpendicular to one of the mirror planes).}
\label{fig:cl_brot}
\end{figure}

One interesting aspect of the FM configuration is that we can examine the behavior of $\sigma_{\text{AHC}}$ as we rotate the magnetization.  For example, we know that the surface AHC wants to align with the magnetization, so when we rotate $\mathbf{M}$ from $[\bar{1}\bar{1}\bar{1}]$ to $[111]$ we should see a sign flip of  $\sigma_{\text{AHC}}$. Of course as $\mathbf{M}$ is rotated, $\sigma_{\text{AHC}}$ has to vanish for some critical angle $\vartheta_c$, and one would naively expect $\vartheta_c=\pi/2$. This is not the generic case, however; nothing forces $\sigma_{\text{AHC}}$ to vanish for $\vartheta=\pi/2$, as can be seen for example in Fig.~\ref{fig:cl_brot}(a), where we find that $\vartheta_c\approx 85^\circ$.  Furthermore, when $\mathbf{M}$ is parallel to the surface, as in Fig.~\ref{fig:cl_brot}(b), $\sigma_{\text{AHC}}$ vanishes only for specific directions.

To understand Fig.~\ref{fig:cl_brot}(b), we turn our attention to the symmetries characterizing the slab. Neglecting $\mathcal{T}$ for the moment, we can obtain the ``slab point group" by considering the subset of the bulk point group that maps the slab onto itself without interchanging the top and bottom surfaces,\citep{taherinejad-prb14} i.e., preserving the $(111)$ direction in the present case.  Note that the pyrochlore point group contains three mirror planes containing the $(111)$ axis, which therefore constitute good symmetries of the slab.  To understand the role of $\mathbf{M}$, we turn our attention to the magnetic point group. $\mathbf{M}$ and $\boldsymbol{\sigma}$ are odd under $\T$ and even under $\I$, so the mirror plane remains a good symmetry only if $\mathbf{M}$ is perpendicular to that mirror plane.  In that case, a non-zero $\sigma_{\text{AHC}}$ is inconsistent with the symmetry, and therefore has to vanish. On the other hand, if $\mathbf{M}$ lies in one of the mirror planes, then the induced symmetry is mirror composed with $\mathcal{T}$ instead of a simple mirror, and this does not reverse  $\sigma_{\text{AHC}}$.  Thus, a non-zero $\sigma_{\text{AHC}}$ is allowed in this case, as well as for generic $\mathbf{M}$ directions for which all mirror symmetries are broken.

\subsection{Termination-dependent surface AHC in the AIAO configuration}
\label{sec:stepchan}

\begin{figure}
\centering\includegraphics[width=\columnwidth]{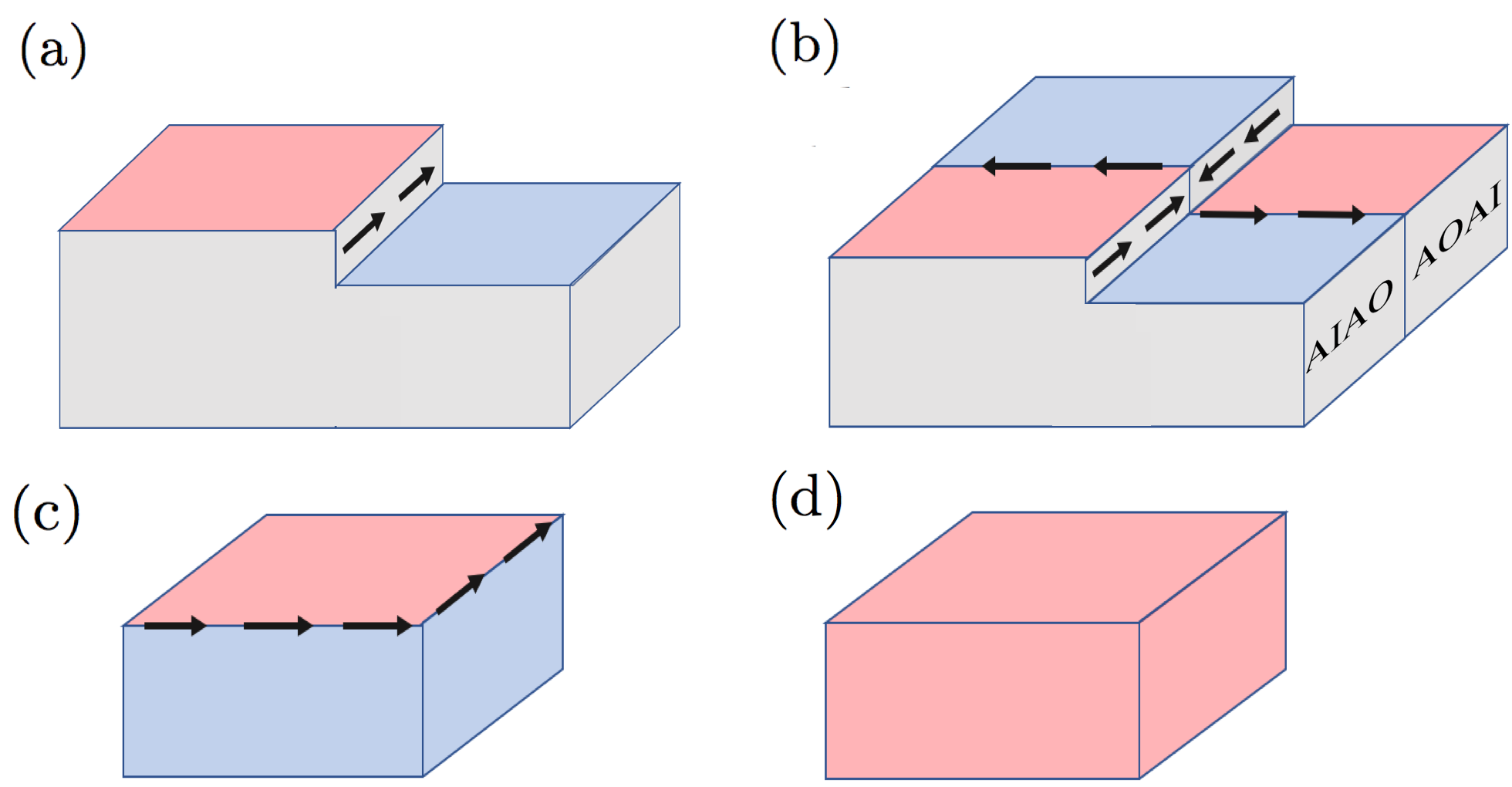}
\caption{Illustration of possible surface AHC configurations for an AIAO insulator in the axion phase with (001)-type surface terminations.  Blue and red colors represent positive and negative surface AHC respectively as defined in an outward-directed convention.  (a) Removal of one atomic layer creates a chiral step channel.  (b) Intersection of a domain wall and a chiral step channel results in a junction with two incoming and two outgoing chiral channels.  (c) Intersection of facets with different signs of surface AHC generates chiral channels along the hinges.  (d) Manipulation of surface terminations by addition or removal of layers can eliminate edge channels and lead to a configuration displaying the topological magnetoelectric effect (see text).}
\label{fig:stepchannel}
\end{figure}

The surface AHC response of an axion insulator in the AIAO spin configuration is fascinating because it is at the same time robust, in the sense that it is half-quantized, and sensitive, since adding or removing a layer flips the sign of the surface AHC.

Fig.~\ref{fig:stepchannel}(a) shows an axion insulator slab where a step has been created by removing a layer (or in general, an odd number of layers) over half the surface. The step is the boundary between two regions with opposite surface AHC, resulting in a chiral boundary mode with conductance $e^2/h$ located at this step. We adopt the convention of coloring the insulating surfaces red or blue according to the sign of the outward-directed surface AHC in this and subsequent figures. This provides easy visual guidance to the location and direction of edge channels, which circulate clockwise around blue facets, i.e., those with positive surface AHC.

The same chiral channel could also be obtained at the intersection of the surface with an antiferromagnetic domain wall (AFM DW), as was first discussed by Mong \textit{et al.}.~\citep{mong-prb10} In Fig.~\ref{fig:stepchannel}(b) an AFM DW separates two regions with AIAO and all-out-all-in (AOAI) spin configurations. Since the AOAI configuration is related to the AIAO by $\T$ symmetry, the resulting surfaces will have opposite AHC. An intersection of an AFM DW channel and a step channel results in a junction with two incoming and two outgoing channels, as in Fig.~\ref{fig:stepchannel}(b). A scattering matrix $T$ should describe how the outgoing amplitudes depend on the incoming ones, with conservation of charge guaranteeing that $T$ is unitary. An STM tip or other object in close proximity with the junction would cause a modification of the unitary scattering matrix, providing possibilities for the construction of a novel quantum switch or sensor.

Another interesting scenario comes into play when we think of macroscopic crystallites. At the edges (or ``hinges'') where facets meet, the facets can have either the same or opposite signs of surface AHC. This is illustrated for the case of a cubic crystallite in Figs.~\ref{fig:stepchannel}(c-d). If one could somehow control the terminating layer at each of the six surfaces, then one could achieve the topological magnetoelectric effect by preparing all the surfaces with the same sign (i.e., color). A crystallite of volume $V$ would then exhibit the full quantized magnetoelectric polarizability of $e^2V/2h$, and would be free of edge channels. In this case we have a rare example of a 3D bulk topological state with no bulk-boundary correspondence at all -- on 2D surfaces, 1D edges, or 0D corners. In the language of ``higher-order TIs,'' this is like saying that the order is higher than the dimension of the bulk.

\subsection{Higher-order topological classification of an FM axion insulators}
\label{sec:hoti}

As we have seen above, the half-quantized surface AHC of an axion insulator may lead to chiral channels flowing along boundaries on its surface.\citep{mong-prb10,sitte-prl12} For example, a macroscopic crystallite will have hinge states at the intersection of two facets with opposite sign of the surface AHC, as was illustrated above. A similar discussion about higher-order states in an inversion-symmetric TI was recently presented in Ref.~\onlinecite{khalaf-prb18}. Here we emphasize how these hinge-state configurations can be regarded as distinct topological phases, such that passing from one to the other, one has to pass through a metallic intermediate phase.  In particular, some facets have to become metallic at such critical points. We will also illustrate how the surface phase diagram can be deduced for such a higher-order topological phase. For this purpose it is most convenient to focus on the case of the FM spin ordering, since this presents the opportunity for easy control via an externally applied magnetic field.

We start with an octahedral crystallite in the FM axion-insulator phase and imagine changing the direction of the magnetization. It is tempting to assume that as long as $\mathbf{M}$ is not parallel to a facet, the perpendicular component of $\mathbf{M}$ will open the surface state gap, with the sign of the half-quantized surface AHC determined by the outward-normal component of $\mathbf{M}$.  By the same token, one may guess that if $\mathbf{M}$ is parallel to a facet, $\sigma_{\text{AHC}}=0$ and the facet will be metallic. While our results are broadly consistent with this picture, we have found that $\sigma_{\text{AHC}}$ does not necessarily pass through zero exactly at the expected 90$^\circ$ critical angle between $\mathbf{M}$ and the surface normal, a fact that we shall return to shortly.  For now, however, we adopt the broad picture in which the direction of $\mathbf{M}$ determines the sign of the surface AHC on each facet in the expected way.

\begin{figure}
\centering\includegraphics[width=\columnwidth]{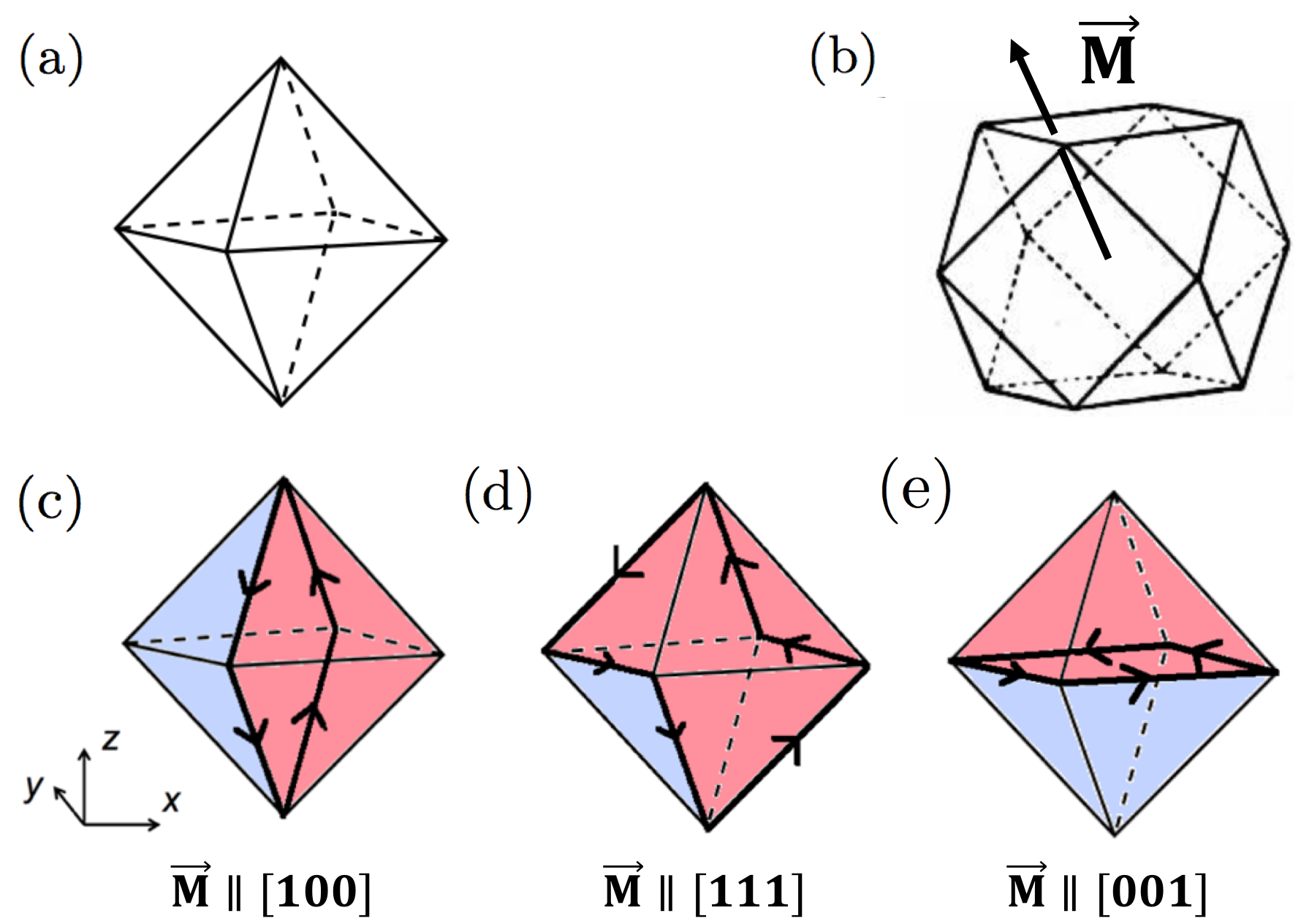}
\caption{ Higher-order topological states at the surface of an FM axion insulator. (a) Octahedral crystallite with facets perpendicular to the (111) and equivalent directions. (b) Higher-order phase diagram. A given magnetization vector intersects the cuboctahedron at a particular face, with each face corresponding to a different arrangement of signs of the surface AHC on the crystallite. (c)-(e) Examples of surface configurations resulting from different magnetization directions. Color scheme follows that of Fig.~\ref{fig:stepchannel}.}
\label{fig:surfacediagram}
\end{figure}

In what follows, we argue that the surface of an axion insulator exhibits multiple higher-order topological phases, and the surface phase diagram is described by a polyhedron embedded in the magnetization sphere, which serves as the parameter space. For example, the surface phase diagram of an octahedral crystallite in real space is described by a cuboctahedron in $\mathbf{M}$ space, as shown in Figs.~\ref{fig:surfacediagram}(a) and (b) respectively, while that of a cubic crystallite is described by an octahedron in $\mathbf{M}$ space. In the former case, it is only when $\mathbf{M}$ intersects an edge of this cuboctahedron that it becomes parallel to one of the crystallite facets, and the surface gap of that facet closes. On the other hand, when $\mathbf{M}$ intersects a face of the cuboctahedron, there is a non-zero perpendicular component of $\mathbf{M}$ on every crystallite facet, resulting in a particular coloring of all facets. When $\mathbf{M}$ is as shown in Fig.~\ref{fig:surfacediagram}(b), for example, the corresponding coloring is that of Fig.~\ref{fig:surfacediagram}(e).

This shows that each face of the cuboctahedron corresponds to a different topological  phase of the crystallite, characterized by a particular chiral loop configuration flowing on the edges of the crystallite.  The edges of the cuboctahedron then correspond to boundaries between two topological phases.

Realization of such a device would enable control of conducting channels using magnetic fields. For example, one can imagine attaching wires to the top, bottom, left, and right corners of the octahedron in Fig.~\ref{fig:surfacediagram}(c-e). Rotating  $\mathbf{M}$ from [100] to [001] will then switch the system between configurations in which the horizontal or vertical wires are connected.  Rotating $\mathbf{M}$ to [111] would instead connect all four wires.  Here we have the potential for yet another kind of novel quantum switch.

We now return to a point mentioned earlier and recall that the picture presented above needs a slight modification.  That is, when tilting the magnetic field orientation, the closure of the surface gap may occur when the magnetic field is near, but not at, the condition of being parallel to the facet, unless this is enforced by some symmetry.  An example was shown in Fig.~\ref{fig:cl_brot}(a), where the critical angle is around 85$^\circ$ instead of 90$^\circ$, while Fig.~\ref{fig:cl_brot}(b) illustrates the role of symmetry.  Taking this effect into account would cause the phase boundaries in Fig.~\ref{fig:surfacediagram}(b) to become slightly distorted, with the edges of the cuboctahedron no longer being perfectly straight (i.e., no longer perfect great circles in $\mathbf{M}$ orientation space). Nevertheless, these distortions must respect the crystal symmetries, and have no effect on the qualitative aspects of the discussion given above. 

\section{Summary and conclusions}
\label{sec:summary}

In this work, we have constructed a minimal tight-binding model of an axion insulator and explored its behavior using a newly developed computational tool that allows for a direct calculations of the surface AHC. This provides us with a virtual laboratory that enables us to answer questions regarding the sign of the surface AHC that could not previously be addressed. We have found that the sign of the surface AHC is extremely sensitive to the surface magnetic configuration, and have shown how the surface AHC is forced to vanish when certain surface symmetries are present.

Our observations distinguish intrinsic axion insulators from magnetically doped strong TIs, which have drawn much attention in recent years.  While the former remain relatively unexplored, they have the advantage of providing exciting opportunites for the control of macroscopic behavior by the manipulation of surface structural and magnetic configurations.  In our pyrochlore model, for example, the pattern of signs of the surface AHC on the crystal facets can be manipulated by a single-atomic-layer surface modification in the AIAO case, or by rotation of a weak external magnetic field in the FM case.  In this way, it may be possible to controllably switch between different configurations of chiral edge and step channels on the surface, or to eliminate such channels altogether in order to achieve the topological magnetoelectric effect.  Such a state is an extreme limit of a higher-order TI in which no protected boundary modes are present.

We nevertheless emphasize that our tight-binding model, though physically motivated, does not describe real insulating pyrochlore iridates with AIAO magnetic order, which instead are believed to lie in the trivial insulator portion of our phase diagram in Fig.~\ref{fig:phase_diagram}. In fact, we are unaware of any experimental demonstration to date of a bulk intrinsic axion insulator. The search for representatives of this intriguing phase of matter presents a pressing challenge for both the theoretical and experimental condensed-matter communities. 
\begin{acknowledgments}
This work was supported by grant NSF DMR-1408838.  We wish to thank Ivo Souza, for useful discussions.
\end{acknowledgments}

\bibliography{pap-used}

\end{document}